\documentclass[aps,prl,reprint,groupedaddress,showpacs,amsmath,amssymb,twocolumn,floatfix]{revtex4-1}
\usepackage{graphicx}
\usepackage{bm}
\usepackage{color}
\usepackage[normalem]{ulem}
\usepackage{cleveref}
\usepackage{amsmath}

\newcommand\redsout{\bgroup\markoverwith{\textcolor{red}{\rule[0.5ex]{2pt}{0.4pt}}}\ULon}

\newcommand{\dlb}{$d/\ell_B$}

\newcommand{\RCF}{$R_{xx}^{CF}$ }
\newcommand{\dragxy}{$R_{xy}^{drag}$ }
\newcommand{\dragxx}{$R_{xx}^{drag}$ }

\begin{document}

\title{Crossover between Strongly-coupled and Weakly-coupled Exciton Superfluids}

\author{Xiaomeng Liu$^{1\dagger}$}
\author{J.I.A. Li$^{2\dagger}$}
\author{Kenji Watanabe$^{3}$}
\author{Takashi Taniguchi$^{4}$}
\author{James Hone$^{5}$}
\author{Bertrand I. Halperin$^{1}$}
\author{Philip Kim$^{1*}$}
\author{Cory R. Dean$^{6*}$}

\affiliation{$\dagger$ X.Liu and J.I.A.Li contributed equally to this work}
\affiliation{$^{1}$ Department of Physics, Harvard University, Cambridge, Massachusetts 02138, USA}
\affiliation{$^{2}$Department of Physics, Brown University, Providence, RI 02912, USA}

\affiliation{$^{3}$Research Center for Functional Materials, National Institute for Materials Science, 1-1 Namiki, Tsukuba 305-0044, Japan}

\affiliation{$^{4}$International Center for Materials Nanoarchitectonics, National Institute for Materials Science,  1-1 Namiki, Tsukuba 305-0044, Japan}

\affiliation{$^{5}$Department of Mechanical Engineering, Columbia University, New York, NY 10027, USA}
\affiliation{$^{6}$Department of Physics, Columbia University, New York, NY 10027, USA}

\maketitle

\textbf{In fermionic systems, superconductivity and superfluidity are enabled through the condensation of fermion pairs. The nature of this condensate can be tuned by varying the pairing strength,  with weak coupling yielding a BCS-like condensate and strong coupling resulting in a BEC-like process. However, demonstration of this cross-over has remained elusive in electronic systems.  Here we study graphene double-layers separated by an atomically thin insulator.  Under applied magnetic field, electrons and holes couple across the barrier to form bound magneto-excitons whose pairing strength can be continuously tuned by varying the effective layer separation.   Using temperature-dependent Coulomb drag and counter-flow current measurements, we demonstrate the capability to tune the magneto-exciton condensate through the entire weak-coupling to strong-coupling phase diagram. Our results establish magneto-exciton condensates in graphene as a model platform to study the crossover between two Bosonic quantum condensate phases in a solid state system.}

\begin{figure*}
\includegraphics[width=0.75\linewidth]{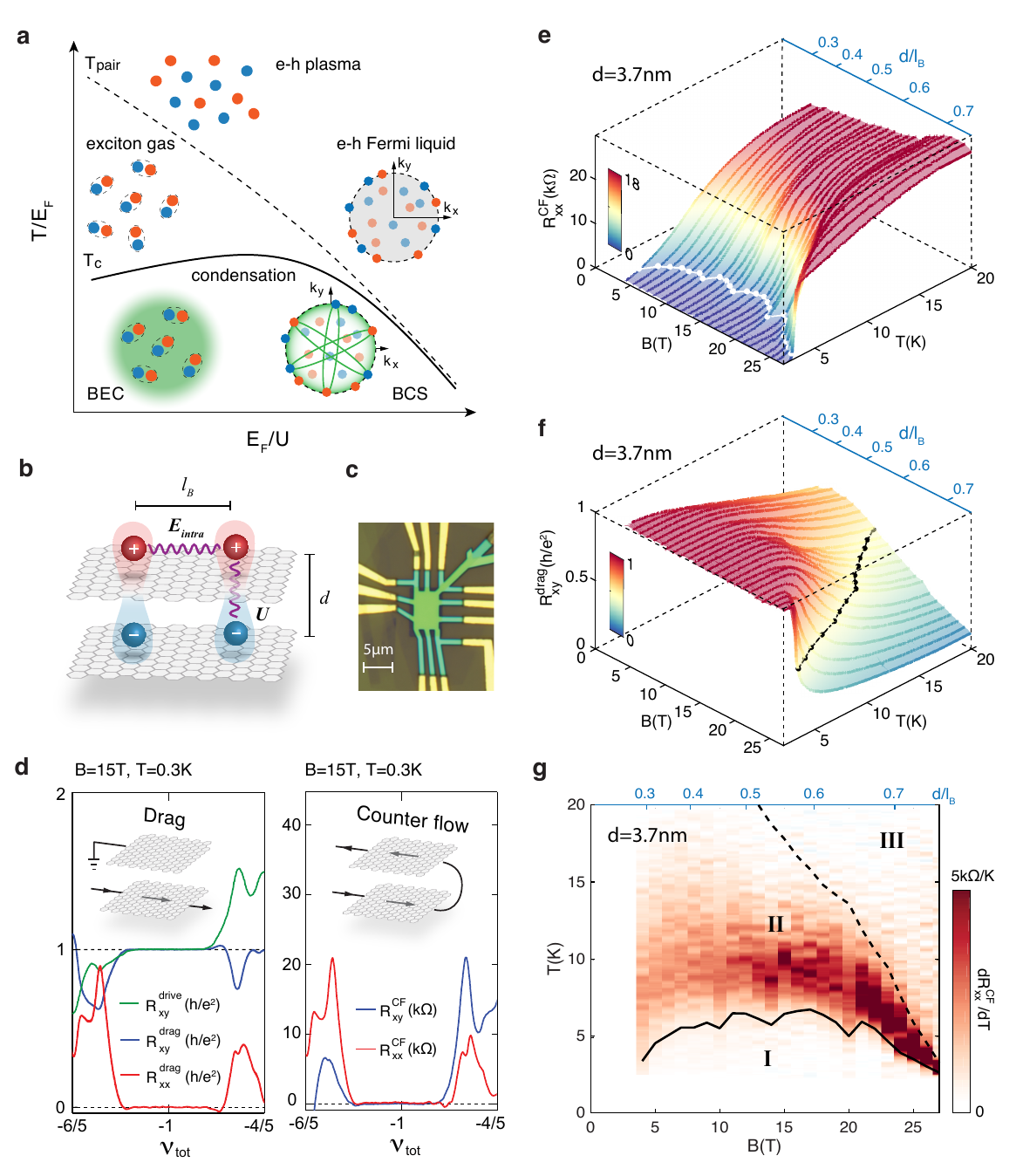}
\caption{\textbf{$\vert$ Two regimes of exciton condensate.} \textbf{a,} Schematic phase diagram for  equal densities  of electrons and holes, with varying temperature and coupling strength. In the strong coupling limit ($E_F/U \ll 1$), the electrons (orange circles) and holes (blue circles) start to pair at $T_{pair}$ and condense at much lower temperature $T_c$. The green halo signifies the condensate. In the weak coupling limit ($E_F/U \gg 1$), the electrons and holes exist as Fermi liquids at high temperatures and establish BCS type of pairing below $T_c$. $k_x$ and $k_y$ are wave-vectors in the x and y direction, while the green lines denote pairing between electrons and holes on the Fermi surface. \textbf{b,} Schematic showing the energy and length scales associated with exciton pairing in a graphene double-layer structure under a magnetic field. Interlayer Coulomb coupling $U$ depends on the interlayer separation $d$, whereas intralayer Coulomb repulsion $E_{intra}$ is determined by the magnetic length $\ell_{B}$. \textbf{c,} Optical image of a graphene double-layer  device used in this study. \textbf{d,} Left panel: Couloub drag response of exciton condensate at $\nu_{tot}=-1$. Inset, schematic for drag measurement setup, arrow indicates the direction of current flow in the drive layer. Right panel: longitudinal and Hall resistance in counterflow geometry measured at $\nu_{tot}=-1$. Inset, schematic for counterflow measurement setup. Arrows indicate the direction of current flow in each layer. \textbf{e,} Waterfall plot of longitudinal resistance from counterflow measurement as a function of temperature measured at different $B$. White line marks the superfluid transition temperature, $T_c$, where $R^{CF}_{xx}$ drop to near zero. \textbf{f,} Waterfall plot of Hall drag response as a function of temperature measured at different $B$.  Black dahsed line marks the pairing temperature $T_{pair}$, where the Hall drag is half of the quantized value. \textbf{g,} Temperature derivative of \RCF\ as a function of temperature $T$ and magnetic field $B$. The black solid and dashed lines mark $T_c$ and $T_{pair}$, respectively, according to the definition in panel \textbf{e} and \textbf{f}. The corresponding \dlb value is marked on the top axis. Area I corresponds to a condensate, area II normal states of excitons and area III normal states of disassociated electrons and holes.
}
\end{figure*}

In the presence of attractive interactions, a fermionic system can become unstable against pairing, forming composite bosons. These paired fermions then can yield a low temperature condensate phase. It has long been recognized that the nature of the fermionic condensate  and its phase transition is directly governed by the strength of the pairing interaction $U$ compared with the Fermi energy $E_F$ as shown in Fig. 1a~\cite{Leggett2012,Randeria2014,Chen2005}. Electrons in metals provide a paradigm example of the weak coupling regime, where pairing interaction is small compared to the Fermi energy ($U \ll E_F$). A low temperature superconducting phase emerges from this weakly interacting Fermi liquid, described by the Bardeen--Cooper--Schrieffer (BCS) theory ~\cite{Bardeen:1957}. In this regime, electrons near the Fermi surface pair in momentum space, with the size of the resulting Cooper pair usually much larger than inter-particle distance ~\cite{Randeria2014}. In the opposite limit of strong interactions ($U \gg E_F$), fermions form spatially tightly bound pairs, and the size of the pair is much smaller than the average inter-particle separation. In this strongly coupled limit, the system behaves like a bosonic gas or liquid, instead of a Fermi liquid, and the low temperature ground state is characterized by a Bose-Einstein condensate (BEC).

A crossover between the BEC and BCS regimes  can theoretically be realized by tuning the ratio of $U/E_F$~\cite{Eagle1969, leggett1980, nozieres1985}, which also corresponds to tuning the ratio of the `size' of the fermion pairs versus the inter-bosonic particle spacing. 
In solid state systems, where the most prominent fermionic condensates, i.e. superconductors,  are found, the BEC-BCS crossover paradigm is highly relevant since, while most metallic superconductors are understood to be in the BCS limit, some unconventional superconductors, such as the high-$T_c$  cuprates ~\cite{Randeria1989,Timusk1999,Tallon2001,Chen2005},  and twisted bilayer graphene~\cite{Cao2018} are thought to reside near the crossover ($U \sim E_F$) between the BEC and BCS limits.  
In cold-fermion gasses, continuous tuning between the weak-coupling and strong-coupling limits has been demonstrated, and the unitary crossover regime firmly established~\cite{Bourdel2004,Regal2004,Bartenstein2004,Zwierlein2004,Ries2015, Murthy2018}. 
Demonstration of this same crossover in a solid state platform (i.e.  within a single electronic superconductor) has not been experimentally realized owing to the inability to continuously tune the coupling strength (e.g. vary $U$ at fixed $E_F$), or the electron density (vary $E_F$ at fixed $U$) sufficiently while maintaining the condensate ground state~\cite{Du2017,Zhu2017} .

In this work, we examine the crossover behavior of the condensate phase of magneto-excitons in quantum Hall bilayer (QHB) systems. Superfluidic magneto-exciton condensation was first realized in QHBs fabricated from GaAs heterostructures~\cite{Eisenstein2014} and subsequently graphene double-layers~\cite{Liu2017,Li2017}. Here, electron-like and hole-like quasi-particles of partially filled Landau levels (LLs) reside in two parallel conducting layers. At integer values of the combined LL filling fraction $\nu_{tot}=\nu_{top}+\nu_{bot}$, where $\nu_{top}$ ($\nu_{bot}$) is the filling fraction of the top (bottom) layer, electrons in one layer and holes in the other layer can pair up, forming interlayer excitons that then condense into a superfluid state at low temperatures~\cite{Eisenstein2014}.

Unlike metallic superconductors, the QHB systems have the advantage that pairing between fermions is widely tunable. Since the kinetic energy of electrons is quenched in the LLs, the energetics of this system is determined by the competition between the intralayer Coulomb interaction $E_{intra}=e^2/\epsilon \ell_B$ (in Gaussian cgs units) where $\ell_B=\sqrt{\hbar/eB}$ is the magnetic length and $\epsilon$ is the background dielectric constant, and the attractive interlayer Coulomb interaction between an isolated electron and hole in the lowest LL, $U\approx (e^2/\epsilon) / ( d + 0.8 \,  \ell_B)$, where $d$ is the interlayer separation (Fig. 1b) [See Supplementary Materials (SM)].
For an isolated layer with a partially filled LL, a Chern-Simons gauge transformation can turn its strongly-interacting electrons with $E_{intra}$ to a composite Fermi liquid with Fermi energy $E_{F} \propto E_{intra}$ \cite{Halperin1993}. In QHBs, the ratio $U/E_{intra}$, which is solely determined by \dlb, therefore provides a characterization of the relative pairing strength, analogous to the dimensionless parameter $U/E_F$ for generic fermionic systems with dispersive bands~\cite{Eisenstein2014,Jerome1967, Littlewood2004}. For $d \ll \ell_B$, $U$ is of the order of $E_F$, resulting in  relatively tightly bound electron--hole pairs, which persist at temperatures well above the transition temperature where the Bose condensate disappears.        
For $d \gg \ell_B$, the two layers are only weakly coupled, with each layer described by a composite Fermi liquid. In this limit, interaction between the two Fermi surfaces can lead to a pairing instability at low temperatures, resulting in a BCS-like condensate. ~\cite{Bonesteel1996,Moller2008,Moller2009,Alicea2009,Sodemann2017}[see SM for more discussion].

\begin{figure*}
\includegraphics[width=1\linewidth]{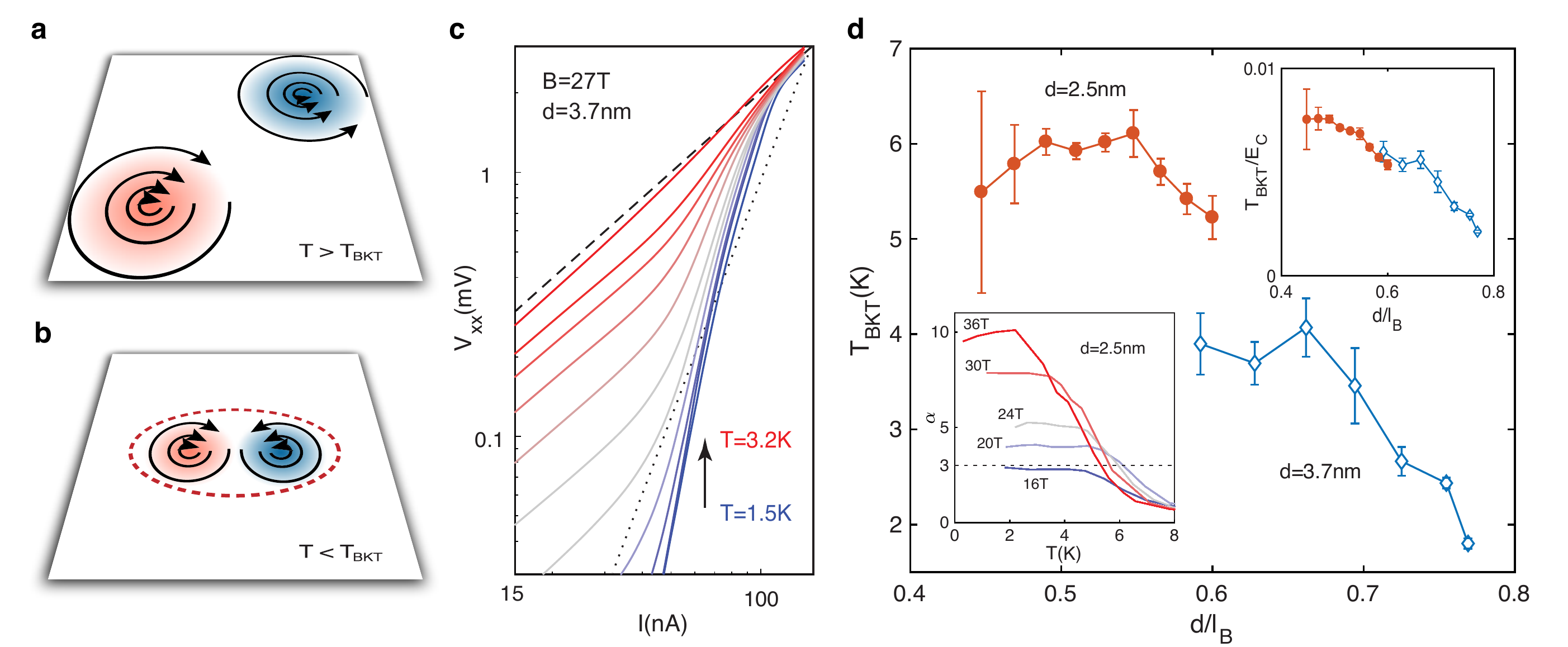}
\caption{\textbf{$\vert$ BKT transition in the BCS regime.} \textbf{a, b,} Illustration of BKT transition. The circling black lines show the winding of the superfluid phase. Blue and red circles represent vortex and anti-vortex. When $T>T_{BKT}$, vortex and anti-vortex are free to move, while below trhe BKT temperature, they are bound into pairs (red dashed line). \textbf{c,} Current-voltage ($IV$) relationship at B = 27T in the d = 3.7nm device at various temperatures. The dashed and dotted lines mark power-law exponent of $\alpha=$1 and 3. \textbf{d,} BKT transition temperature defined as power $\alpha=3$ as a function of $d/l_B$ in two samples with different interlayer separation. Bottom left inset: $\alpha$ extracted from the $IV$ curve as a function of temperature. Under high magnetic fields, $\alpha$ rises above three at low temperatures, as expected for a BKT transition. However, the value of $\alpha$ saturates at low temperatures, and as the magnetic field drops, the saturation value decreases. Eventually, for smaller magnetic fields, $T_{BKT}$ cannot be defined,  as  $\alpha$ saturates below three. (See, for example,  B = 16T in Fig. 3d bottom inset). Top right inset: BKT transition temperature after scaling to Coulomb energy $E_C$. Data from two samples with different interlayer separation collapse onto a universal line. 
}
\end{figure*}
 
Experimentally, $d/\ell_B$ can be continuously varied in a single device, by varying the applied magnetic field $B$, or across multiple devices, by changing the interlayer distance $d$. This provides the opportunity to continuously tune through the complete condensate phase diagram.  In our study, we fabricated QHBs from graphene double-layers, consisting of two parallel graphene layers separated by a few layer hBN dielectric tunneling barrier (Fig.1b).  We focus on the magneto-exciton condensate appearing at $\nu_{tot}=-1$, corresponding to both layers filled to half filling of the first hole LL ($\nu_{top}=\nu_{bot}=-1/2$), although similar behavior was observed for other integer values of $\nu_{tot}$. We report results over the range $0.3<$\dlb $< 0.8$, where well-defined exciton superfluid states exist at the lowest experimental temperature.

To probe the dynamics of the interlayer exciton we utilize the Coulomb drag and counterflow (CF) geometries~\cite{Su2008,Kellogg2002,Kellogg2004,Tutuc2004a} (shown schematically in the inset of Fig. 1d.  [See SM]. In the Coulomb drag geometry, the exciton condensate is identified by the emergence of a quantized Hall resistance plateau equal to $h/e^2$, as measured in both the drive and drag layer, concomitant with zero longitudinal resistance on both layers (Fig. 1d). In contrast, when the two layers are decoupled, the drive layer exhibits the density-dependent Hall resistance, while the Hall resistance of the drag layer is close to zero~\cite{Liu2017D}. Thus, the Hall drag resistance $R_{xy}^{drag}$ provides an experimental measure of interlayer pairing ~\cite{Eisenstein2014,Liu2017, Li2017}. In the counterflow geometry, charge neutral excitons can be induced to flow by configuring the current to move in opposite directions in the two layers\cite{Eisenstein2004}.  In this geometry the neutral exciton current gives a zero valued Hall resistance in both layers, while the dissipationless nature of the superfluid condensate is revealed by a vanishing longitudinal resistance (Fig. 1d). 
 
Figures~1e \& f show the temperature dependence of the counterflow longitudinal resistance \RCF and Hall drag resistance  \dragxy of a d=3.7nm device, for different values of \dlb  which is tuned by varying the magnetic field $B$ (also see Fig.~S1). At low temperatures, the exciton superfluid phase is observed over the full range of effective layer separation that we studied, $0.3<$ \dlb $<0.8$, evidenced by the vanishing \RCF\ in CF and quantized \dragxy ~\cite{Eisenstein2014,Kellogg2002,Kellogg2004,Tutuc2004a}.  

The temperature evolution of these quantities across different $d/ \ell_B$  allows us to experimentally map key features of the condensate phase diagram.  First we identify the critical temperature of the condensate as the value below which the  longitudinal resistance becomes dissipationless. Practically we define this as the temperature where \RCF drops to less than 5\% of the high temperature saturation value. Indicated by a white line in Fig. 1e, this boundary identifies a dome below which the condensate is well formed. The dome shape of the critical temperature is consistent with theoretical expectation \cite{Littlewood2004}. In the strong coupling limit (small \dlb), the primary consequence of increasing B is a corresponding increase of the exciton density ($\propto$ B) which in turn drives up $T_c$. Oppositely, in the weak coupling limit (large \dlb), increasing \dlb\   further reduces the interlayer coupling, resulting in a diminishing of the pairing between the two Fermi liquids and causing $T_c$ to decrease.

Second, we interpret \dragxy\ as a measure of the pair fraction. In the limit of strong coupling, where electrons and holes occur in tightly-bound pairs, excitons may persist at temperatures well above the counterflow-superconductivity critical temperature. In this temperature range, we would still expect to observe a large \dragxy\  response. On the other hand, at temperatures high enough such that electrons and holes are dissociated, the value of \dragxy\ will be close to zero. 
We can therefore identify a temperature scale for the pair-breaking  by the temperature  where \dragxy\ deviates significantly from the quantized value $h/e^2$. Phenomelogically, we define the pair-breaking temperature, $T_{pair}$, as the temperature where \dragxy\ drops to half its quantized value, i.e. $h/2e^2$ (indicated by a black line in Fig. 1f).

In Fig. 1g we summarize the experimental phase diagram by plotting the temperature derivative of the counterflow resistance, $dR^{CF}_{xx}/dT$, versus \dlb.  Plotting this way emphasizes the three distinct regimes of the magneto-excitons phase diagram: the low temperature superfluidic condensate (Phase I, $T<T_c$), the intermediate phase where there is a dissipative channel, i.e. $R^{CF}_{xx}>0$, but the two layers remain coupled through exciton formation (Phase II, $T_c<T<T_{pair}$); and the high temperature normal phase where the layers are decoupled and most excitons are unbound (Phase III, $T>T_{pair}$). We note that the temperature range over which $dR^{CF}_{xx}/dT$ is finite valued  tracks reasonably well the $T_c$ and $T_{pair}$ phase boundaries identified from Figs. 1e and 1f, respectively, indicating $R^{CF}_{xx}$ and $R^{drag}_{xy}$ are correlated in this phase diagram and dissipation continuously increases with temperature in Phase II.

The experimental phase diagram shown in Fig. 1g additionally reveals distinct temperature behaviour between the small \dlb~(strong coupling) and large \dlb~(weak coupling) regimes. At small \dlb, $T_{pair}$ is much larger than $T_c$,  with a gradual transition observed between the condensate phase (Phase I)  and the high-temperature layer-decoupled phase (Phase III). This signifies that in the strong coupling limit the exciton pairing establishes well above the condensation temperature, consistent with the behavior expected for a BEC condensate. In contrast, at large {\dlb},  $T_{pair} \sim T_c$, indicating that in the weak coupling regime interlayer pairing and fermion pair condensation occurs simultaneously, which is the very signature of BCS superconductors. The similarity of these behaviors at small and large \dlb\ to the well-known temperature dependence of the BEC and BCS pictures (depicted in Fig. 1a), establish graphene double-layer as a uniquely tunable platform where fermion pair condensation can be studied in both strong and weak pairing regime ~\cite{Randeria2014,Leggett2012,Chen2005,Littlewood2004}.

\begin{figure*}
\includegraphics[width=0.95\linewidth]{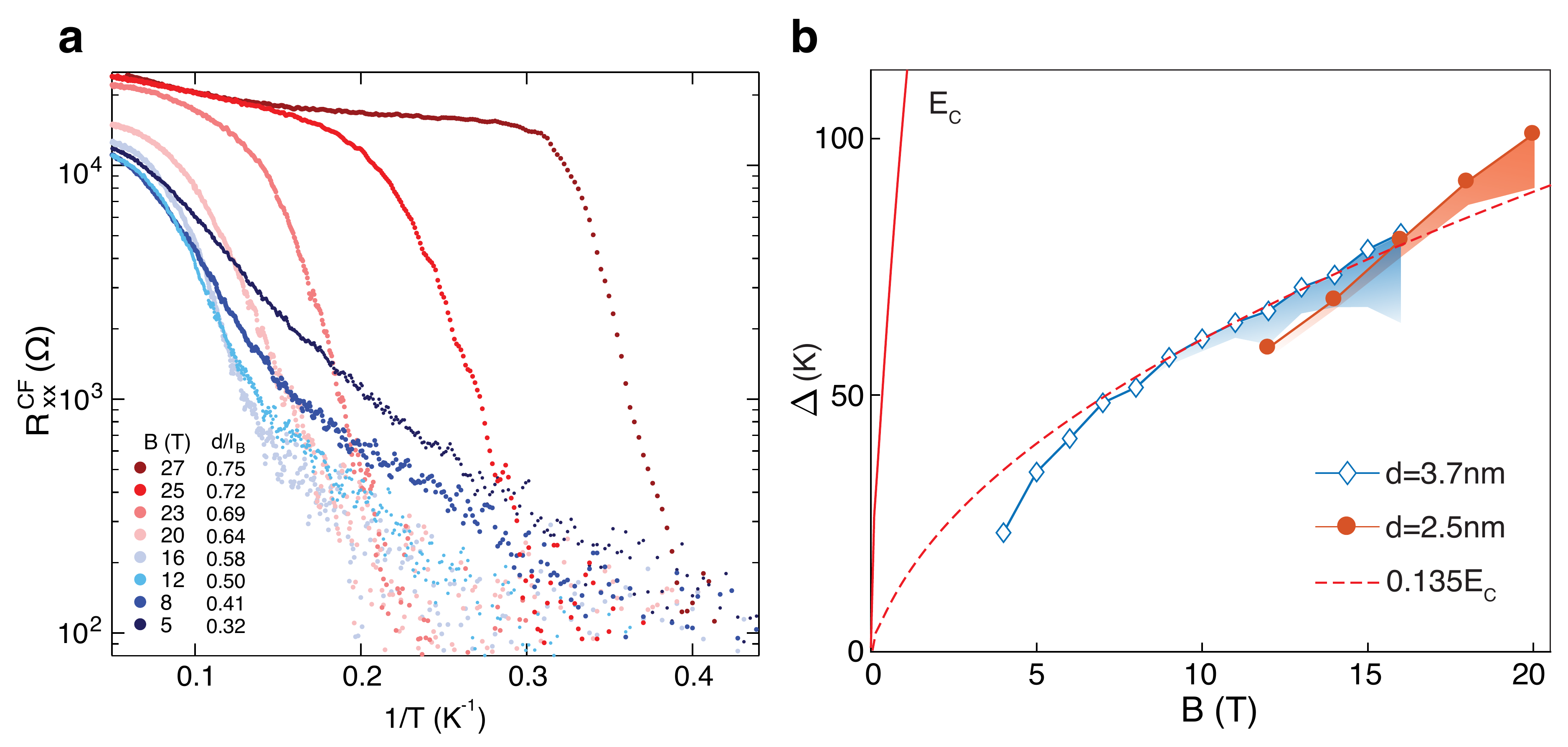}
\caption{\textbf{$\vert$ Activation energy in the strong coupling regime.} \textbf{a,} Arrhenius plot of \RCF\ measured at different magnetic fields in the d=3.7nm device. \textbf{b,} Activation gap $\Delta$ as a function of magnetic field for two devices with different interlayer separation $d = 3.7$~nm and $2.5$~nm. Red solid curve correspond to Coulomb energy, $E_c = e^2/\epsilon\ell_{B}$, where $e$ is electron charge and $\epsilon$ the dielectric constant of hBN. Red dashed curve labels $0.135E_c$. }
\end{figure*}

The condensate phase transitions of magneto-excitons in QHBs can be further examined in the context of its 2D nature. At $T<T_c$, the exciton condensate is expected to be a 2D superfluid described by the Berezinskii-Kosterlitz-Thouless (BKT) theory ~\cite{berezinskii1972,KT1973,Girvin2000}. In order to produce a counterflow voltage, it is necessary that topological defects, namely vortices in the condensate order parameter (Fig. 2a,b),  should move across the sample in a direction perpendicular to the voltage gradient. Since the energy of an isolated vortex in a 2D superfluid diverges logarithmically with the size of the system, vortices can exist at low temperatures only in bound pairs of opposite signs (Fig. 2b). Counterflow resistance would not be produced by motion of such pairs. As temperature rises, the vortices unbind at the critical temperature $T_{BKT}$ (Fig. 2a). Above $T_{BKT}$, the movement of free vortices leads to a counterflow resistance. Below $T_{BKT}$, although the linear counterflow resistance is predicted to vanish, there can be a non-linear response, giving a non-zero voltage at finite measuring currents. Specifically, it is predicted that for small currents $I$, one should find a power law relation: $V \propto I^{\alpha}$, where the exponent is given by $\alpha=1+4 \rho_s (T)/\pi T $, where $\rho_s(T)$ is the temperature-dependent  phase-stiffness constant for the order-parameter.  
According to BKT theory, $T_{BKT} = \frac{\pi}{2} {\rho_s}(T_{BKT})$, so $\alpha$
should be equal to  3  at $T_{BKT}$ and should increase monotonically with decreasing temperature below that\cite{Halperin1979}.       
In principle, the measured exponent should jump discontinuously to $\alpha=1$ above $T_{BKT}$, but this decrease should  only be gradual for finite measuring current.

Fig.~2c plots experimental $IV$ curves measured in the counterflow geometry in logarithmic scale. 
For our smallest measuring currents,  below $\approx 100$~nA, we indeed observe power-law behavior, and we extract a measured exponent
$\alpha (T)$, by fitting the slope of the $IV$ curve at low currents. The result is plotted as a function of $T$ in the bottom left inset of Fig.~2d. At large \dlb, $\alpha$ increases with decreasing $T$, allowing us to extract  $T_{BKT}$ based on the criterion of $\alpha=3$. Fig.~2d shows the experimentally-obtained $T_{BKT}$ over a large range of \dlb\ for two graphene double-layer devices. In the large \dlb limit, $T_{BKT}$ obtained from the $IV$ curves follows that of the critical temperature $T_c$ in Fig. 1g. 

In the BCS framework, $\rho_s(T)$ collapses at the mean-field transition temperature $T_m$ due to the proliferation of unpaired quasiparticles and thus $T_{BKT}$ is bounded by the mean-field transition temperature $T_m$~\cite{Halperin1979}. Since increasing \dlb\ corresponds to weakening the interlayer BCS pairing, $T_m$, and thus $T_{BKT}$, should decline as \dlb\ increases, in agreement with the experimental observation shown in Fig 2d for \dlb$>0.5$.
As \dlb\ decreases from the BCS limit, we find that $T_{BKT}$ first increases, and then tends to saturate as the \dlb\ reaches $\sim$0.5, following the trend of $T_c$. However, eventually the BKT transition becomes ill defined.  
As can be seen in the inset of Fig.~2d, even for large magnetic fields, the measured value of $\alpha$ does not diverge as predicted for $T \to 0$, but  saturates at a finite value. The saturation  value decreases with decreasing $B$, and eventually falls below 
 $3$.  The mechanism behind low-temperature saturation of $\alpha$ is unclear, but may relate to the gradual evolution of counterflow resistance as a function of temperature at small \dlb, including possible effects of disorder. Interestingly, we find that $T_{BKT}$ measured from two samples collapses  onto a universal curve after scaling with Coulomb energy, $E_c=e^2/\epsilon \ell_B$, as shown in the right inset of Fig.~2d, demonstrating the critical role of Coulomb interaction underlying the emergence of exciton condensate in graphene double-layers. 

As $B$ decreases, we move from the BCS limit (high $B$) to the BEC limit (low $B$), we find the transition to low temperature condensation phase changes qualitatively. Fig. 3a shows an Arrhenius plot of $R^{CF}_{xx}$ versus temperature at fixed values of the applied magnetic field, $B$. Whereas a sharp jump in $R^{CF}_{xx}(T)$ is seen at large {\dlb}, consistent with the BKT transition described above,   at small {\dlb} the counterflow resistance exhibits a thermally activated behavior,  $R^{CF}_{xx}(T) \sim e^{-\Delta/2T}$ with a well defined $\Delta$ (blue traces in Fig. 3a). 

Plotting $\Delta$ as a function of $B$ in the small \dlb regime
provides insight into the relevant low energy excitations in the BEC limit (Fig.~3b). For both samples, the plots are well fit by  $\Delta = 0.135 E_c$. We note that this value is  significantly  smaller than the energy to create a free electron and hole, indicating that the appearance of the finite resistance is not due to unbinding of excitons. The most relevant collective excitations  in the small \dlb\ limit are predicted to be merons and anti-merons \cite{Girvin}, which are charged topological vortices of the exciton condensate, with large core radii. [See SM]  Merons have core energies that are a fraction of $E_c$ and it may be argued that in the extreme limit of \dlb$\to 0$, there may be a regime where  the density of free merons leads to $R^{CF}_{xx} \sim e^{- \Delta / 2T}$, with $\Delta$  a fraction of $E_c$. Our estimation of $\Delta$ for generation of a pair of meron-anti meron is $\sim 0.6 E_c$ (see SM), much larger than the observed  $\Delta$, suggesting that disorder might play a crucial role.  
 
We  note that a similar activated behavior of the counter flow current has been observed in GaAs QHBs, but in much weaker coupling parameter range \dlb=1.3 -- 1.8 ~\cite{Kellogg2002,Kellogg2004,Tutuc2004a,Lay1994}. While the cause of the difference between the phenomenology of the two systems is yet uncertain, the atomically thin interlayer separation of graphene QHBs makes our system less susceptible to the influence of disorder, by providing two orders of magnitude larger activation gaps. 

Overall, our results demonstrate that the adjustable pairing strength in graphene double-layer structures allows access to two distinct regimes of fermion pair condensation, characterized by strong and weak coupling strength, where we uncovered distinct transport behaviors and roles of topological excitations. This dynamical and continuous tunability of fermion pairing in a solid-state device opens the door to investigate phenomenology of fermion condensate of various pairing strength, paving the way for improved understanding of the connection between the BCS-BEC crossover and unconventional superconductivity.

\begin{acknowledgments}

\noindent
\textbf{Acknowledgements} C.R.D acknowledges partial support by the US Department of Energy (DOE), Office of Science, Basic Energy Sciences (BES), under award number DE-SC0019481. X.L. acknowledges support  by the DOE (DE-SC0012260) for device fabrication and measurement. PK acknowledges the DoD Vannevar Bush Faculty Fellowship N00014-18-1-2877. Sample preparation at Harvard was supported by ARO MURI (W911NF-14-1-0247). Sample fabrication at Columbia University was supported the Center for Precision Assembly of Superstratic and Superatomic Solids, a Materials Science and Engineering Research Center (MRSEC) through NSF grant DMR-1420634. The theoretical analysis was supported in part by the Science and Technology Center for Integrated Quantum Materials, NSF Grant No. DMR-1231319. K.W. and T.T. acknowledge support from the Elemental Strategy Initiative conducted by the MEXT, Japan, A3 Foresight by JSPS and the CREST (JPMJCR15F3), JST. A portion of this work was performed at the National High Magnetic Field Laboratory, which is supported by the National Science Foundation Cooperative Agreement No. DMR-1644779 and the state of Florida. Nanofabrication  at the Center for Nanoscale Systems at Harvard  was supported  an NSF NNIN award ECS-00335765.    We thank S. H. Simon, B. Lian, S. D. Sarma, I. Sodemann, I. Kimchi, M. Shayegan, and J. P. Eisenstein for helpful discussion. 

\end{acknowledgments}

\noindent
*Correspondence should be addressed to: 
\noindent
pkim@physics.harvard.edu; cdean@phys.columbia.edu.

\bibliography{BECBCS.bib}

\section{Supplementary Materials}

\subsection{Sample fabrication and measurement}
The graphene double-layer devices are made of stacks of hBN-graphite-hBN-graphene-thin hBN-graphene-hBN-graphite (from top to bottom), which are prepared by mechanical exfoliation and the van der Waals transfer technique. The dual graphite gates shield the graphene layers from impurities and contaminations. Separate contacts are fabricated on each individual layer. In Fig.~1c, the top and the bottom leads  contact the bottom graphene layer, and left and right leads contact the top graphene layer. No tunneling current is detected between the two layers. At last, contact gates fabricated above an atomic layer deposited Al$_2$O$_3$ are used to improve the contact transparency of the top graphene layer, while the silicon back gate is utilized to improve the contact transparency of the bottom graphene layer.

Measurements are conducted using AC lock-in technique with 2nA excitation current at 17.7Hz. Coulomb drag measurement is performed by flowing current $I$ in the drive layer while the other layer (the drag layer) is open circuit, with one contact connected to the ground through a 1$M\Omega$ resistor to allow gating. Longitudinal and Hall voltages are measured simultaneously on both graphene layers. Hall resistance on the drive layer is defined as $R_{xy}^{drive} = V_{xy}^{drive}/I$, whereas longitudinal and Hall resistance on the drag layer are defined as $R_{xx}^{drag} = V_{xx}^{drag}/I$ and $R_{xy}^{drag} = V_{xy}^{drag}/I$, respectively. Unless otherwise specified, all data presented in this work are taken at filling factors $\nu_{top}=\nu_{bottom}=$-1/2, which is realized by adjusting the top and bottom gate voltages at each magnetic field.

Counterflow measurements are performed by flowing currents of the same magnitude $I$ in the opposite directions in the two layers. Longitudinal and Hall voltages: $V_{xx}^{CF}$ and $V_{xy}^{CF}$ are measured on the top graphene layer and converted to counterflow resistances: $R_{xx,xy}^{CF} = V_{xx,xy}^{CF}/I$. $IV$ measurements are performed with the AC+DC technique, where a 2nA AC current plus a DC current is passed through the sample. This results in the differential resistance presented in Fig.~S3. $IV$ curves are then obtained by integrating the differential resistance.

\subsection{Additional Data}
Fig.~S1 shows counterflow resistance and drag resistance in both the longitudinal and the transverse direction under a few representative magnetic fields in the 3.7nm sample. We note that the temperature dependence of the longitudinal counterflow resistance is very similar to that of the Hall counterflow resistance. On the other hand, the drag resistance behaves quite differently between the longitudinal and transverse components. As temperature rises, the Hall drag resistance decreases monotonically, signifying the two layers become increasingly decoupled. For longitudinal drag, it first increases as the sample exits the superfluid phase, but decreases when the two layers become decoupled.

Fig.~S2 shows the drag resistance in the longitudinal and transverse direction in the 2.5nm sample. The longitudinal drag resistance is used to extract the activation gap shown in Fig. 3b in this sample.

Fig.~S3 compares the differential resistance behaviors in the BCS regime and the crossover regime in the 3.7nm sample. In the BCS regime (B=27T), the longitudinal counterflow differential resistance shows a clear critical current behavior, similar to that of a superconductor. In the crossover regime (B=13T), however, the differential resistance increases smoothly. Fitting this $IV$ curve with a power law, the power $\alpha$ saturated below three even for the lowest measuring temperature, rendering the BKT temperature ill-defined. 

\subsection{Theory for larger values of \dlb.}

In the limit of large \dlb, our system becomes two separated monolayers at $\nu=1/2$. Because of the applied magnetic field, one cannot describe the isolated layer  simply as a Fermi sea of weakly-interacting electrons.  However, by making a unitary transformation, the single layer system  can be described as a Fermi sea of ``composite fermions", interacting with a Chern-Simons gauge field as well as the applied magnetic field, in such a way that the average net magnetic field felt by the fermions is zero.\cite{Halperin1993}   Then, at the mean-field level,  the separated two-layer system can be described as consisting of a pair  Fermi seas of composite fermions with identical Fermi wave vectors, $k_F = \ell_B^{-1}$.

At finite separations $d$, one must consider the effects of interactions between the layers. As for a a conventional metal with two spin states, the pair of Fermi seas can be unstable to  formation of pairs between the two layers, for arbitrarily weak interactions of appropriate type, leading to a BCS-type ground state.\cite{Bonesteel1996}  The physical properties of the resulting state will depend on the particular form of the pair wave function favored by the interaction, and it is not known what will be the favored form at large separations.  However, as was shown by M\"oller, Simon, and Rezayi, if the composite fermions pair in a state with $p_x + i p_y$ symmetry, the ground state will exhibit the same macroscopic properties as the exciton condensate state obtained at small separations.\cite{Moller2008,Moller2009} Specifically, the state will have an energy gap for deviations of the total filling factor from the value $\nu_{total} = 1$, but will have no energy gap for transferring electrons from one layer to the other. It will behave like a superfluid for counterflowing currents in the two layers, and it will have a Goldstone mode with a linear spectrum at long wavelengths, which is the superfluid sound mode, or the magnon mode in the pseudospin description.  M\"oller {\it{et al.}} explored trial wave functions based on the $p_x+ip_y$ pairing description and found satisfactory overlap with wave functions obtained from exact diagonalizations of systems containing  a small number of particles at intermediate layer separations. 

An alternative description was proposed by Sodemann {\it{et al.}},\cite{Sodemann2017} based on the formulation of the $\nu=1/2$ monolayer in terms of Dirac composite fermions, introduced by D. T. Son.\cite{Son2015} In this case, the pairing wave function compatible with an exciton condensate has $s$-wave symmetry. However, 
there has been no proposal of a trial wave function derived from this formulation to describe either the single-layer or the double-layer system. 

It is not intuitively  obvious in either of the  above descriptions how the system converts in the strong coupling limit to a system of tightly-bound excitons, with $s$-wave-paired electrons and holes. For this reason, we find it useful to employ  another alternative description.\cite{Halperin2020}  Instead of attaching Chern-Simons flux quanta to the electrons in each layer, we attach the  flux to electrons in one layer and holes in the other.  We then form a state with  $s$-wave pairing between the electron-like composite fermions in one layer and the hole-like composite fermions in the other.  

 This approach  uses the fact that for electrons with no internal  degrees of freedom confined to a single LL, with purely two-body interactions,  there is an exact particle-hole transformation, which preserves all energy eigenstates and eigenvalues, except for an additive constant that depends on the total particle number. Furthermore, there is an exact mapping between the energy states of a double-layer system (a) with electrons at filling $\nu$ in the top layer and filling $1- \nu$ in the bottom layer and those of another system (b)  in which the top  layer has filling $\nu$ of electrons while the bottom layer contains positively charged particles at the same density, {\it{i.e.,}} a system  that  is overall neutral.  Furthermore the transport properties of the two systems, at any given temperature, will be related by a simple transformation,  Specifically, if we apply electric fields $\mathbf{E}^{top}$ and $\mathbf{E}^{bottom}$ to the two layers, the electric currents induced in the two systems will be identical, except that the currents in the two bottom layers  will differ by an amount $(e^2/h) \hat{z} \times \mathbf{E}^{bottom}$. 

We now consider a system with electrons at $\nu=1/2$ in the top layer and an equal number of holes in the bottom layer. Let $z_j$ and $w_k$  denote the positions of the electrons and holes respectively, in complex notation, and let $\Psi[\{ z_j \}, \{w_k \}]$ be the wave function describing the system at some time $t$. We introduce a unitary transformation  such that $\Psi$ is related to the transformed wave function $\Psi'$ by
\begin{equation}
\label{transform}
\Psi = \Psi'  \prod_{j<j'} \left[  \frac {z^*_j - z^*_{j'}} {|z^*_i - z^*_{j'} |} \right]^2
\prod_{k<k'} \left[ \frac {w_k - w_{k'}} {| w_k - w_{k'}|} \right] ^2 .
\end{equation}
After implementing the  corresponding unitary transformation on the Hamiltonian, we find that the Hamiltonian that governs the time-dependence of $\Psi'$ contains an induced Chern Simons field that cancels the applied magnetic field, on  average, for the transformed fermions in each layer.  Thus, if we ignore interactions between the fermions, we find a mean-field ground state in which $\Psi'$ describes a Fermi sea of composite fermions in each layer. With interactions between the layers, this may be replaced by a BCS-type ground state:
\begin{equation}
\Psi' = \prod_{\mathbf{k}} [u_{\mathbf{k}} + v_{\mathbf{k}} c^{\dagger} _{\mathbf{k}} 
d^{\dagger} _{- \mathbf{k}} ] \, | 0 \rangle ,
\end{equation}
where $c^{\dagger} _{\mathbf{k}}$ and $d^{\dagger} _{\mathbf{k}} $ are creation operators for a fermion with momentum ${\mathbf{k}}$ in the top and bottom layers respectively,  
$u _{\mathbf{k}}$  and $v _{\mathbf{k}}$ are variational parameters with 
$u _{\mathbf{k}} = [1 - | v _{\mathbf{k}} |^2 ]^{1/2} $, and $| 0 \rangle$ is the vacuum state.  We assume $s$-wave pairing, so that $v_{\mathbf{k}}$  depends only on the magnitude of $ \mathbf{k} $.

Neither the  BCS state nor the unpaired mean-field wave function for separated layers will produce a good wave function $\Psi$ when substituted into (\ref{transform}),  because $\Psi$ will contain a  large admixture of electrons and holes in the higher LLs.  Better wave functions are obtained by projecting the right-hand side of (\ref{transform}) onto the lowest LL. Small system numerics implementing  wave functions related to these  \cite{Wagner-inprep} have found that they perform roughly as well as the p-wave composite fermion wave functions of \cite{Moller2008}.

 Now we can see how the system behaves in two limits. We obtain the limit  of two uncoupled  layers by setting  $u _{\mathbf{k}} = 0$, $v _{\mathbf{k}}= 1$ for $k < k_F$, and $u _{\mathbf{k}}= 1, \,  v _{\mathbf{k}}=0$  for $k> k_F$. On the other hand, we can obtain the limit of tightly bound excitons by letting $v _{\mathbf{k}}$ be a constant, independent of $k$ up to a very large ultraviolet cutoff. (The value of this constant should be chosen to give the desired  total density of particles and  will depend on the value of the cutoff.) In this limit, the composite fermions will only occur in pairs that are tightly bound in position space, so the corresponding values of $z$ and $w$ are identical.  Therefore, the product of phase factors in (\ref{transform}) will be equal to unity. Before projection onto the lowest LL, $\Psi$ will be the wave function for a Bose condensate of non-interacting small-radius excitons. After projection onto the lowest LL, the excitons will have radii of order $\ell_B$. It can be shown that the resulting wave function $\Psi$  is the exact ground state for the double-layer system with separation $d=0$, if mixing between LLs can be neglected. 
 
 This procedure can be easily generalized to the case where the original electron system has $\nu_{total}=1$ but unequal populations of the two layers.  The corresponding electron-hole system will still have equal populations of electrons and holes, but with $\nu \neq 1/2$ in each layer. Then, the effective magnetic fields seen by the composite fermions will be different from zero, and the mean-field eigenstates of the fermions,  for well separated layers, will be simply  states in the LLs  of the effective magnetic field in each layer.   The states for composite fermions in the two layers will be related by complex conjugation, so one can form them into Cooper pairs in the usual manner. Thus a gapped BCS paired state can exist over a continuous range of fillings, as long as the electron and hole fillings are equal.  Of course, if $\nu$ is a  rational fraction for which an isolated layer has a strong fractional quantized Hall state, the existence of an  energy gap in that state will mean that BCS pairing can only exist if the interlayer interaction strength exceeds a certain threshold, implying that the  layers must not be too far apart.      Numerical evaluations of such imbalanced wave functions for small systems give support to this picture.\cite{Wagner-inprep}

\subsection{Estimation of the interlayer binding $U$ }

We define $U$ as the binding energy of an isolated exciton in a state where $\nu_{total} =1$.  Thus, we may consider a situation where there is only one exciton in the system.   Suppose that all electrons except one are in the bottom layer, so we have one electron in the top layer and one hole in the bottom layer.  The exciton binding energy is the difference in energy between the situation where the electron and hole are very far apart and the situation where they are close together,   forming their lowest-energy bound state. 

The wave function for an electron-hole pair in the lowest LL is uniquely determined by its total momentum $\mathbf{k}$. The mean value of the in-plane separation $\mathbf{s}$ between the positive and negative charges is given by $\langle \mathbf{s} \rangle = \hat{z} \times \mathbf{k} \ell_B^2$. Consequently, for an attractive interaction between the electron and hole, the minimum energy of the exciton will occur at $\mathbf{k}=0$. At this wave vector, the correlation function between the electron and hole is given by
\begin{equation}
g(\mathbf{s})=\frac {1}{2 \pi \ell_B^2} e^{- s^2 / 2 \ell_B^2} .
\end{equation} 
The binding energy of the exciton is then given by 
\begin{equation}
U = \int d^2 \mathbf{s} \, g(\mathbf(s)  v_{int} (s) ,
\end{equation} 
where $- v_{int}$ is the attractive interaction between a pair of point charges of opposite sign in the two layers..  For the case of interest, where $v_{int} (s) = e^2 / [\epsilon (s^2 +d^2)^{1/2} ]$, one finds.
\begin{equation}
\label{integral}
U = (e^2/\epsilon) \ell_B^{-2}  e^{d^2 / 2 \ell_B^2} \int_d^{\infty} ds \, e^{-s^2 / 2 \ell_B^2} .
\end{equation}
This right-hand side of (\ref{integral}) reduces to $ (\pi/2)^{1/2} (e^2 / \epsilon \ell_B)$ for $d=0$ and approaches $e^2/(\epsilon d)$ for $d \gg \ell_B$. 
The simple   approximation  quoted in the main text,  $U \approx (e^2/\epsilon) (d+  0.8 \ell_B)^{-1} $, interpolates between these two limits. 

\subsection{Theory of dissipation in the $d/l_B \to 0$ limit}

In this section, we discuss theory of $\nu_{tot}=-1$ state in the limit of $d/l_B \to 0$, with both layers equally populated. Assuming there is no disorder nor interlayer tunneling, the ground state of the double layer system for non-zero \dlb\  can be characterized by an XY-like quantum Hall pseudo-spin ferromagnet, with the pseudo-spin pointing in an arbitrary direction in the XY plane, for equal population of the two graphene layers. (More generally, the z-component of the  pseudo-spin can be nonzero, reflecting the difference in occupation of the top and bottom  layers.) The superfluid phase below $T_{BKT}$ is characterized by quasi-long-range order of the pseudo-spin order parameter, meaning that the correlation function for components in the XY plane falls off as a power law at large distances.

If we assume a pure Coulomb interaction between the electrons and we neglect Landau-level mixing, and if we also assume $d/l_B \ll 1$, we obtain for the phase stiffness of the pseudo-spin at $T=0$,  \cite{KallinH84,SondhiKKR93}  
\begin{equation}
    \rho_s^0 =\frac{1}{\sqrt{512\pi}}(\frac{e^2}{\epsilon l_B}).
\end{equation}
In the limit of $d/l_B = 0$, the XY pseudo-spin ferromagnet acquires  SU(2) symmetry (upgraded from the U(1) symmetry of XY magnet) and becomes a Heisenberg-like pseudo-spin ferromagnet. In this case, there is no quasi-long-range order at any finite temperature.   The phase stiffness constant $\rho_s(T)$ is suppressed to zero at any finite temperature due to pseudo-spin fluctuations in the  z-direction, so  $T_{BKT}$ is driven to 0.  Correlations for the XY components of the pseudosppin will decay exponentially with distance, with a correlation length that diverges as
\cite{BrezinZ76}
\begin{equation}
    \xi \approx l_B \, \exp(2\pi \rho_s^0  /T).
\end{equation} 

For any nonzero \dlb, however, pseudo-spin fluctuations in the  z-direction cost extra energy, so that $T_{BKT}$  may be reduced but will  be nonzero.  In the limit of small but finite \dlb, it is estimated  that  $T_{BKT} \sim  4 T^0_c \,  |\ln (d/\ell_B) |^{-1} $, where $T^0_c = \pi \rho_s^0$ is the bare   BKT transition temperature. However, this estimate is clearly  inapplicable for realistic values of \dlb , which are much greater than $e^{-4}$. 

For small nonzero \dlb, vortices have spread out cores and are known as {\em{merons}}.  Inside the core, the  pseudo-spins tilt into  the $\pm z$ direction, as illustrated in Fig.~S4. 
There are four types of merons, which have positive or negative vorticity and  carry electrical charge $\pm$e/2, concentrated  in one or the other of the   layers.

If the thermally excited merons are the only excitation in the system, and if there are no impurities, then we would expect both the counterflow resistivity and the total longitudinal conductivity to be proportional to the density of unpaired merons. In the limit of very small \dlb, for temperatures above $T_{BKT}$ but still small compared to $T^0_{BKT}$, the density of thermally excited merons is expected to be on the order of
\begin{equation}
    \xi^{-2} \approx l_B^{-2} \,  \exp(-4\pi \rho_s^0 /T).
\end{equation} 
In this case, thermal activation gap $\Delta$ would be predicted to be $8\pi \rho^0_s \sim 0.62 e^2/\epsilon l_B$, which is more than four times  the value obtained in our experiments.   Moreover,   it is probable   that the experimentally relevant values of \dlb\ are  not small enough for this estimate to be applicable, and it is not clear that there will even be a regime of activated resistance in the absence disorder.

If impurities are taken into account, or if the total density is not precisely $\nu_{tot}$ = 1, then there may be a finite density of merons even at T = 0. These will be localized at T = 0, but they should be able to move at finite temperature. Then the activation energy for conductivity may be controlled by an activation energy for motion, which might be substantially  lower than  $8\pi \rho^0_s$.

\renewcommand\figurename{Figure S}
\setcounter{figure}{0}    

\begin{figure*}
\includegraphics[width=\linewidth]{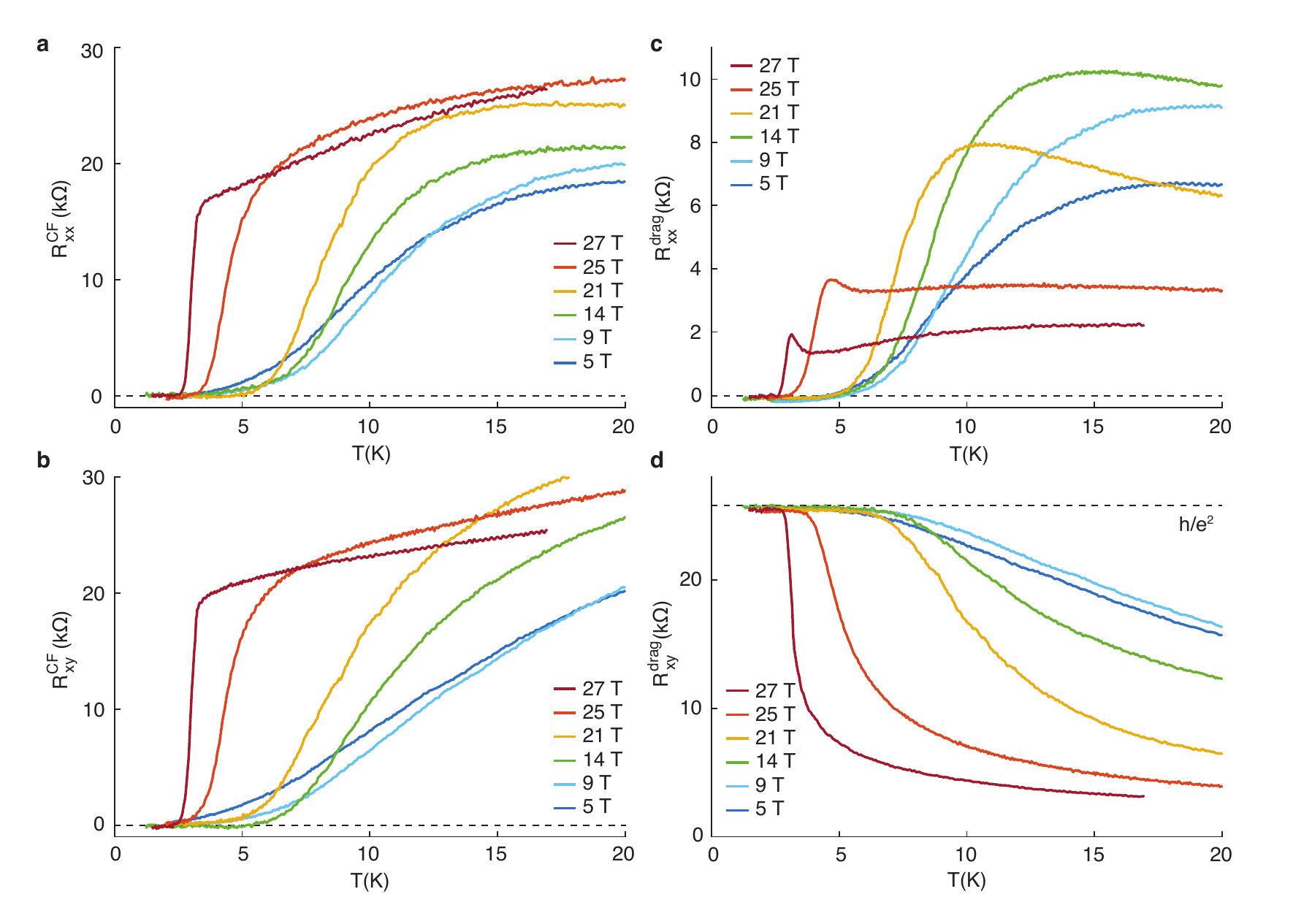}
\caption{\textbf{Temperature-dependent resistance behaviors in the 3.7nm device.} \textbf{a, b,} Counterflow resistance in the longitudinal and transverse direction as a function of temperature under various magnetic fields. \textbf{c, d,} Longitudinal and Hall drag resistance as a function of temperature under the same set of magnetic fields.. }
\end{figure*}

\begin{figure*}
\includegraphics[width=\linewidth]{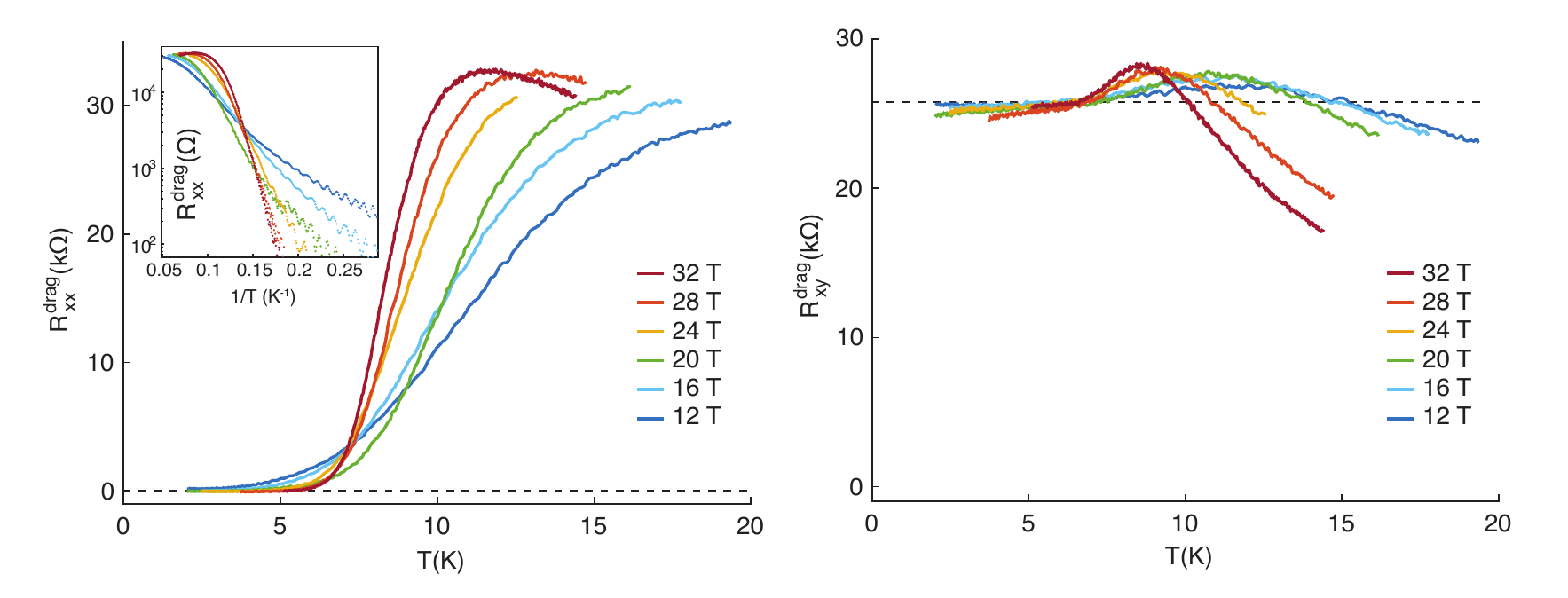}
\caption{\textbf{Drag resistance behaviors in the 2.5nm device.} Temperature dependence of \dragxx (left panel) and \dragxy (right panel) measured at different magnetic fields. }
\end{figure*}

\begin{figure*}
\includegraphics[width=\linewidth]{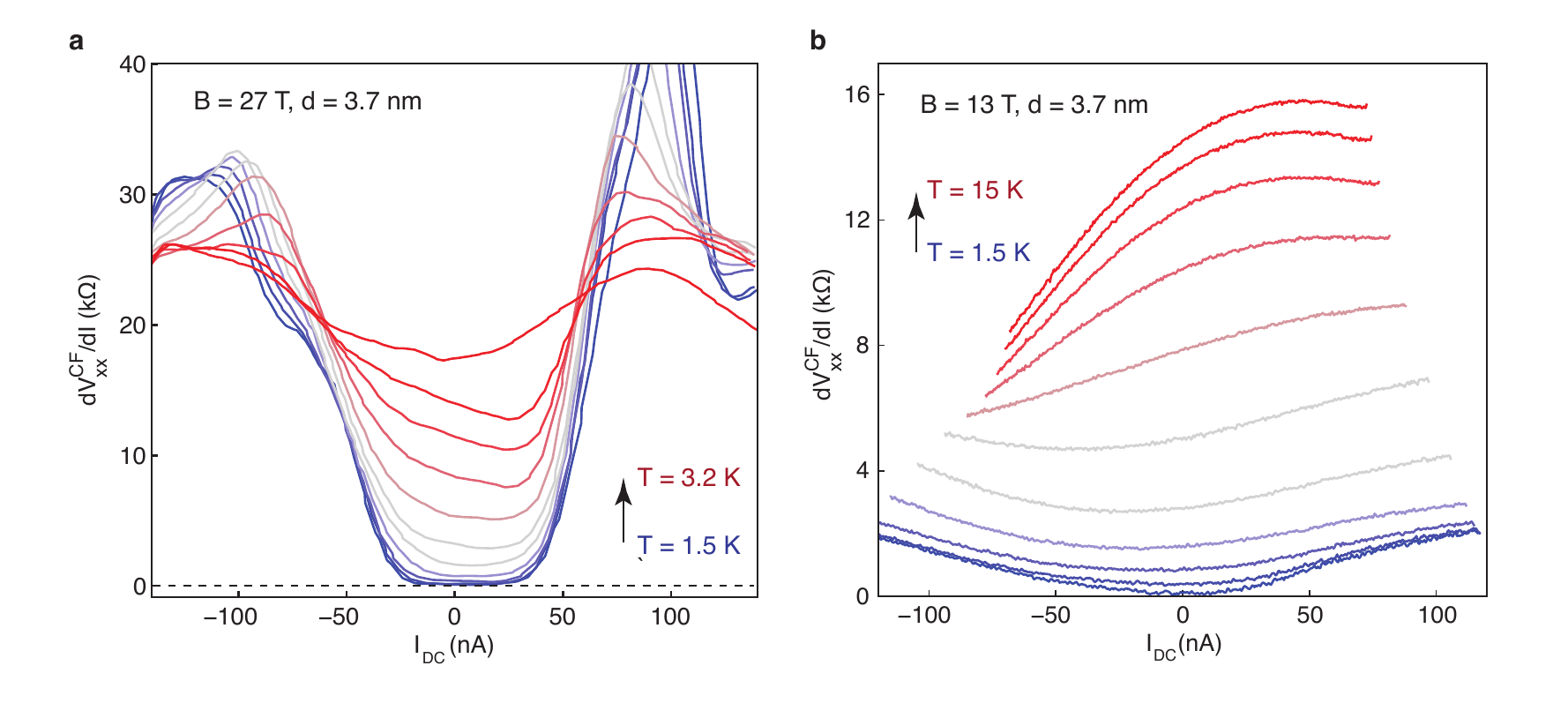}
\caption{\textbf{Different counterflow resistance in different regimes.} \textbf{a,} Differential counterflow resistance in the 3.7nm device under magnetic field of 27T. A clear critical current behavior is noted under low temperatures. \textbf{b,} Differential counterflow resistance under magnetic field of 13T. In contrast to \textbf{a}, there is no clear critical current.}
\end{figure*}

\begin{figure*}
\includegraphics[width=0.5\linewidth]{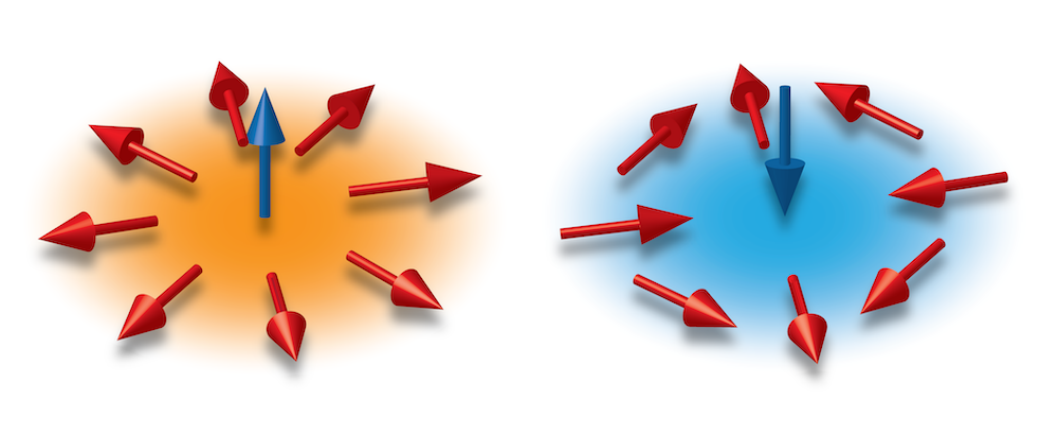}
\caption{\textbf{Illustration of meron and anti-meron.}  The figure shows the orientation of the pseudo-spin as a function of position in the plane of the sample. Up and down arrows show regions where electrons are in the upper or lower layers, respectively, while horizontal arrows represent coherent equal  admixtures of the  two states.  The illustrated configurations have opposite vorticity, but carry the same electric charge. Configurations with the opposite charge are obtained by reversing the directions of the blue arrows.  }
\end{figure*}

\end{document}